\documentclass{article}
\usepackage[utf8]{inputenc}
\usepackage[margin=1in]{geometry}
\usepackage{subcaption}
\usepackage{graphicx}
\usepackage{amsmath}
\usepackage[hyphens]{url}
\usepackage{hyperref}
\hypersetup{colorlinks=false,breaklinks=true,hidelinks}
\usepackage{cleveref}
\usepackage{apacite}
\usepackage{natbib}
\usepackage{authblk}
\usepackage{multirow}
\usepackage{booktabs}
\usepackage{makecell}
\graphicspath{{figures/}}
\usepackage{setspace}
\title{Towards End-to-End Earthquake Monitoring Using a Multitask Deep Learning Model}
\author[1,2]{Weiqiang Zhu\thanks{zhuwq@berkeley.edu}}
\author[1,2]{Junhao Song}
\author[1,2]{Haoyu Wang}
\author[3]{Jannes Münchmeyer}
\affil[1]{Department of Earth \& Planetary Science, University of California, Berkeley}
\affil[2]{Berkeley Seismology Laboratory, University of California Berkeley}
\affil[3]{Université Grenoble Alpes, Université Savoie Mont Blanc, CNRS, IRD, Université Gustave Eiffel, ISTerre, Grenoble, France}

\date{}

\begin{document}

\maketitle

\begin{abstract}

    Seismic waveforms contain rich information about earthquake processes, making effective data analysis crucial for earthquake monitoring, source characterization, and seismic hazard assessment. 
    With rapid developments in deep learning, the state-of-the-art approach in artificial intelligence, many neural network models have been developed to enhance earthquake monitoring tasks, such as earthquake detection, phase picking, and phase association. However, most current efforts focus on developing separate models for each specific task, leaving the potential of an end-to-end framework relatively unexplored.
    To address this gap, we extend an existing phase picking model, PhaseNet, to create a multitask framework. This extended model, PhaseNet+, simultaneously performs phase arrival-time picking, first-motion polarity determination, and phase association. The outputs from these perception-based models can then be processed by specialized physics-based algorithms to accurately determine earthquake location and focal mechanism. The multitask approach is not limited to the PhaseNet model and can be applied to other state-of-the-art phase picking models, ultimately improving seismic monitoring through a more unified and efficient approach.
    
\end{abstract}

\section{Introduction}

Earthquake catalogs, which comprehensively record earthquake frequency, location, magnitude, and mechanisms, are fundamental for characterizing earthquake sources, studying earthquake physics, analyzing underlying geological processes, and enhancing earthquake forecasting and hazard mitigation.
Recent advancements in earthquake monitoring focus on improving catalog completeness by developing innovative algorithms to distinguish weak seismic signals from background noise, enabling the detection of smaller earthquakes. 
As the number of earthquakes increases exponentially with decreasing magnitude, more sensitive methods often lead to 5 to 10 times larger catalogs than previous methods.
These catalogs thereby provide detailed insights into fine-scaled earthquake source processes, such as stress transfers or fluid migrations.

Deep learning, the state-of-the-art machine learning approach in artificial intelligence (AI), is increasingly utilized to detect small earthquakes from seismic archives due to its high detection sensitivity and processing efficiency.
A significant advantage of deep learning models compared to conventional algorithms such as STA/LTA (short-term averaging/long-term averaging) \citep{allen1978automatic} is that deep neural networks can automatically learn effective features from labeled data to achieve detection accuracy comparable to human experts.
Deep learning models have been widely applied to detect various types of seismic events and build enhanced earthquake catalogs, including tectonic earthquakes \citep{tan2021machine,liu2020rapid,liu2023complexity,neves2024complex}, induced earthquakes \citep{park2020machine,chai2020using,park2022basement,wang2020injection}, volcanic earthquakes \citep{wilding2023magmatic,retailleau2022automatic,zhong2024deep}, subduction zone earthquakes \citep{xi2024deep,jiang2022detailed}, low-frequency earthquakes \citep{thomas2021identification,munchmeyer2024deep,lin2023detection}, and laboratory earthquakes \citep{shi2024labquakes}.

However, most existing efforts focus on developing separate deep learning models for individual tasks, such as phase picking, first-motion polarity picking, phase association, and event location.
For example, widely adopted deep learning models, such as GPD \citep{ross2018generalized}, PhaseNet \citep{zhu2019phasenet}, and EQTransformer \citep{mousavi2020earthquake}, are specifically designed for detecting and picking seismic phase arrivals.
These models are trained to classify each sample within a 30-60 second window of seismic waveform data into one of three categories: P-phase, S-phase, or noise.
They achieve both high accuracy in phase detection and high resolution in arrival time determination \citep{munchmeyer2022which}.
Compared to phase arrival-time picking models, fewer models exist for phase polarity picking \citep{ross2018p,zhang2023simultaneous,zhao2023ditingmotion,wang2025seismollm,han2025rpnet,peng2025microseismic}.
Estimating first-motion polarities is an essential step for estimating focal mechanisms, with a more sensitive polarity detector unlocking focal mechanisms even for small events.
Current polarity pickers follow an approach similar to \citet{ross2018p}'s model: given a segment ($\sim$4s) of waveform centered on the P wave arrival, the model classifies the arrival into three categories: ``Up", ``Down", or ``Unknown".
For phase association, deep learning models are still in early development stages, with examples including PhaseLink based on recurrent neural networks \citep{ross2019phaselink} and GENIE based on graph neural networks \citep{mcbrearty2023earthquake}. Commonly used phase association models rely on conventional methods, such as grid-search-based techniques \citep{zhang2019rapid,munchmeyer2024pyocto}, or alternative machine learning approaches, such as RANSAC and Gaussian mixture models \citep{woollam2020hex,zhu2022earthquake}.

Earthquake monitoring workflows combining deep learning models with conventional source parameter inversion algorithms \citep{klein2002user,kissling1994initial,waldhauser2001hypodd,trugman2017growclust,hardebeck2002new,skoumal2024skhash} have proven efficient for detecting earthquakes and studying complex earthquake sequences \citep{zhu2023quakeflow,zhang2022loc}.
However, two limitations exist in the current sequential processing workflow: cumulative information loss across steps and inefficient processing of large data archives. For example, phase detection and picking processes only identify signals above a certain threshold, overlooking coherent but weak signals below this threshold; Phase association only processes the picked phase arrivals without considering phase waveform coherence; Phase polarity picking requires prior phase arrival detection to extract short waveform windows for polarity determination, introducing additional complexity and computational overhead—particularly burdensome for processing large seismic archives. Consequently, deep learning-enhanced catalogs, which have become routine for studying earthquake sequences, often lack focal mechanism solutions.
Current developments primarily focus on enhancing separate deep learning models for each task, leaving the potential for an integrated, end-to-end framework largely unexplored.

In this work, we aim to address these challenges by developing a multitask model capable of simultaneously performing phase arrival time determination, polarity picking, and earthquake origin times estimation.
To this end, we introduce PhaseNet+, a model that jointly estimates picks, polarities, and origin times with a single forward path.
While phase arrival times and polarities are essential for describing the earthquake source, the estimation of origin times greatly simplifies phase association.
As a multitask model, PhaseNet+ uses a unified hidden representation of the waveforms for all tasks, leveraging mutual information between them.
This enhances the overall model performance compared to separately training each task, as many key features for the tasks are shared.
For example, both P phase arrival picking and polarity picking will require knowledge about the relevant frequency bands, or the understanding of the first signals corresponding to an arriving wave.
Moreover, as this unified approach extracts all necessary information, including phase arrivals, polarities, and origin times, from a single scan of the continuous data, the processing times are substantially accelerated in comparison to individual models.
The improved performance and simplified processing facilitate the construction of comprehensive earthquake catalogs with both high-resolution location and focal mechanism information.

\section{Method}

\subsection{Multitask deep learning architecture}
\begin{figure}
    \centering
    \includegraphics[width=0.9\textwidth]{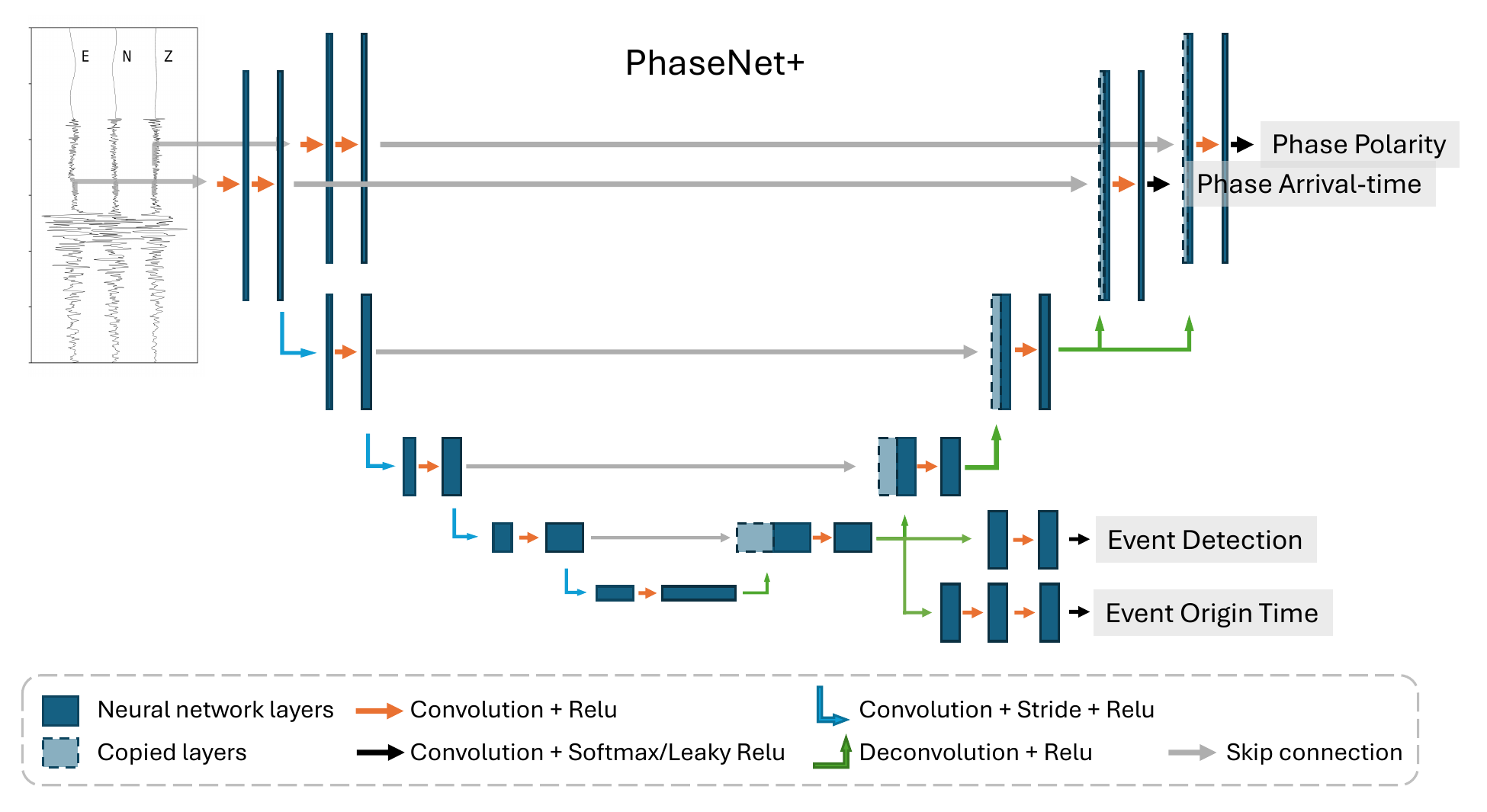}
    \caption{Neural network architecture of the PhaseNet+ designed for simultaneous phase arrival-time picking, phase polarity picking, and origin time prediction. The phase arrival-time branch processes three-component input data, while the phase polarity branch processes only the vertical component. Encoded features from the shared deep layers are also fed into the polarity branch (indicated by the green arrow), guiding the model to select relevant waveforms for phase polarity picking. Event detection and origin time prediction are conducted in deeper layers, where a broader view of seismic waveforms including both P and S arrivals is necessary and high temporal precision is less critical. The predicted source parameters are subsequently used for phase association.}
    \label{fig:model_architecture}
\end{figure}

To enhance earthquake monitoring accuracy and efficiency through an end-to-end workflow, we developed PhaseNet+, a multitask model that simultaneously performs phase arrival-time picking, phase polarity picking, and earthquake origin time prediction for phase association (\Cref{fig:model_architecture}).
\paragraph{Phase arrival-time picking}
Deep learning models have demonstrated robust performance in phase arrival-time picking tasks \citep{munchmeyer2022which}. Our multitask deep learning model builds upon the PhaseNet architecture \citep{zhu2019phasenet}, which approaches phase picking as a semantic segmentation task using a U-Net architecture \citep{ronneberger2015unet}. PhaseNet classifies each sample of a seismic waveform into one of three categories: P-phase, S-phase, or noise. This approach enables jointly performing phase detection, i.e., identifying if there is a seismic arrival at all, and phase picking, i.e., estimating the precise temporal time of the phase arrival.
In contrast to, e.g., a regression setup, this approach makes no assumptions on the number of arrivals in a target window and can thereby handle even multiple arrivals.
In addition, the sequence-to-sequence approach allows for efficient processing of continuous data without requiring sliding windows.
As the label for arrival-time picking, we employ a truncated Gaussian function, similar to PhaseNet.
\paragraph{Phase polarity picking}
Traditional phase polarity picking models typically treat the task as a binary classification problem.
For each P pick, a small segment of the surrounding waveforms is categorized as either ``Up" or ``Down", sometimes with an additional ``Unknown" category \citep{ross2018p}.
This classification approach introduces an inherent dependency of the polarity picking on the accuracy of the P pick and the appropriate window selection.
Inaccurate initial P-wave detection can lead to erroneous polarity reversals due to missing critical waveform features, resulting in unstable predictions \citep{zhao2023ditingmotion}.
Our model incorporates phase polarity detection into the semantic segmentation task. We extend the original PhaseNet architecture with a new convolutional layer branch designed to learn corresponding features for phase polarity picking, generating prediction scores between [-1, 1] at each sample. The polarity and arrival-time branch share the same hidden layers, up to the last deconvolution.
The polarity scores can be directly extracted at the sample identified by phase arrival-time branch.
This multitask design ensures a shared hidden representation, allowing for accurate phase arrival detection and classification while enabling simultaneous polarity determination in continuous seismic data. 
\paragraph{Phase association}
Phase association traditionally involves identifying coherent phase picks following moveout patterns of travel-times and amplitudes. While basic approaches like grid search can accurately identify potential earthquakes by examining possible origin times and locations, they are computationally intensive. We explore an alternative approach to phase association through direct prediction of seismic event origin times.
Origin time can be estimated from single station records, for example, through the travel-time difference between P and S phases, a classical method for manually estimating origin time or epicentral distance \citep{shearer2019introduction}. 
We integrate this principle into our multitask model by adding a dedicated branch to the PhaseNet architecture for event detection and origin time prediction. This branch performs two functions: detecting the average of P and S arrival-times as a segmentation task and predicting event origin times as a regression task (\Cref{fig:model_architecture}). Using this approach, phase association simplifies to identifying picks with shared onset times, substantially simplifying the problem. While the first function may introduce occasional ambiguities when different events exhibit similar midpoints of P and S arrival times, it is acceptable considering efficiency.

\begin{figure}
    \centering
    \includegraphics[width=0.8\linewidth]{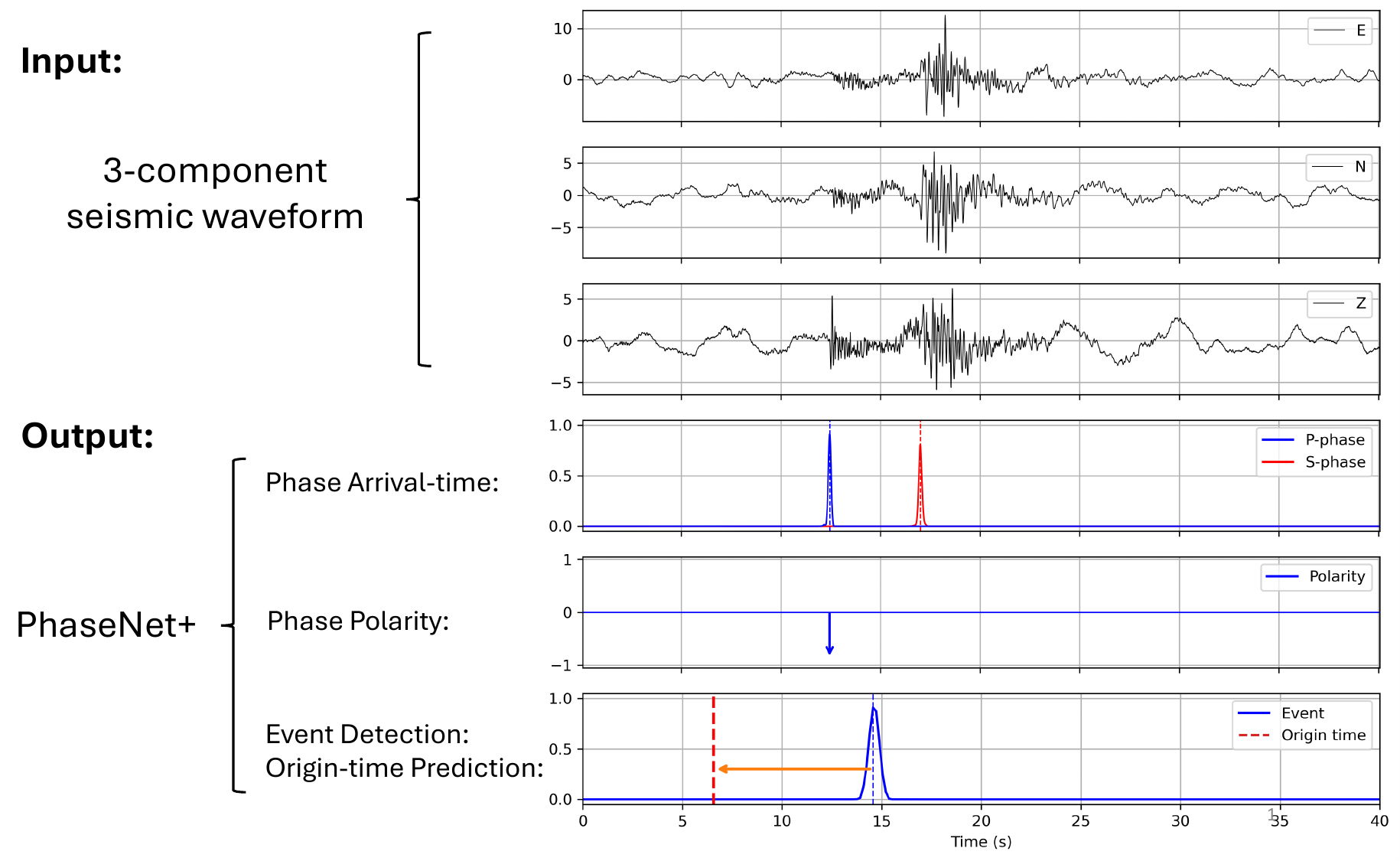}
    \caption{Example of PhaseNet+ predictions demonstrating the model's multitask capabilities. The model processes input waveforms to simultaneously generate three types of information: phase arrivals (P and S phases), phase polarities (Up and Down), and event detection with origin-time predictions (indicated by dashed orange lines).}
    \label{fig:example}
\end{figure}

The multitask model architecture integrates phase arrival-time picking, phase polarity picking, and event detection with origin time prediction, as illustrated in \Cref{fig:model_architecture}. This architectural framework is adaptable beyond PhaseNet+; all other deep learning approaches modeling phase picking as a sequence-to-sequence task, such as EQTransformer \citep{mousavi2020earthquake} and PhaseNO \citep{sun2023phase}, can be extended to multitask implementations following similar principles. \Cref{fig:example} demonstrates PhaseNet+ predictions, showing how the model processes three-component seismic waveforms to generate comprehensive output including phase arrivals, polarities, and event detection with origin-time predictions.

\subsection{End-to-end earthquake monitoring workflow}

\begin{figure}
    \centering
    \includegraphics[width=\textwidth]{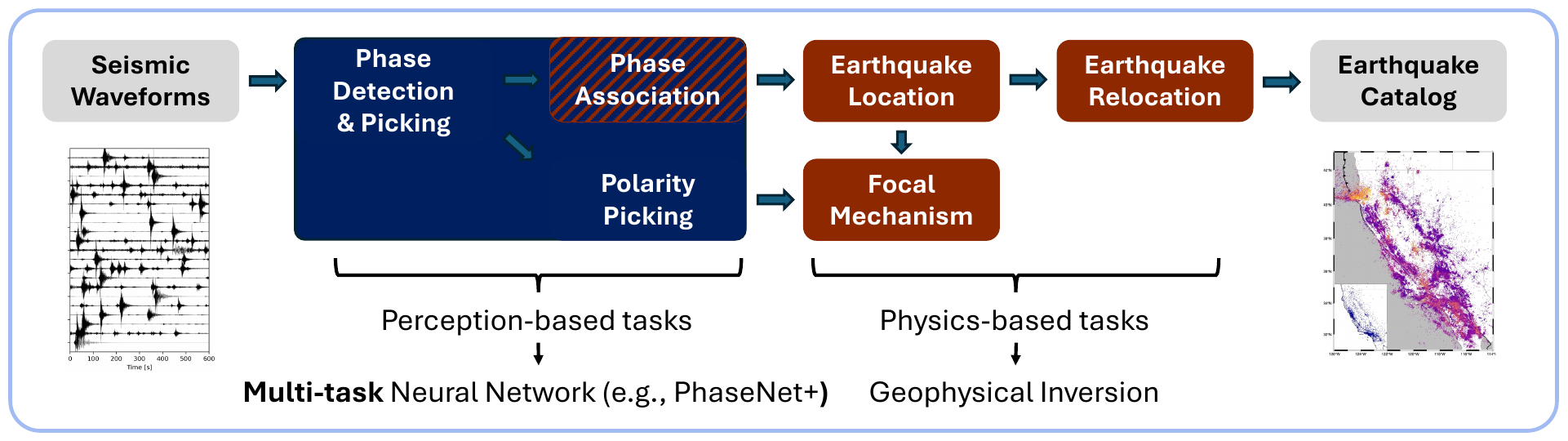}
    \caption{Overview of the end-to-end approach to earthquake monitoring based on a multi-task model. The model is designed to solve the perception-based tasks of phase arrival-time detection/picking and polarity picking and provide constraints (i.e., origin time prediction) for phase association, all through a single scan of continuous waveforms. The model outputs are then directly utilized for determining earthquake locations and focal mechanisms, enabling the construction of comprehensive earthquake catalogs.}
    \label{fig:workflow}
\end{figure}

Earthquake monitoring tasks can be divided into two categories: perception-based and physics-based tasks (\Cref{fig:workflow}). Perception-based tasks, such as phase arrival-time picking and phase polarity picking, focus on extracting information from seismic waveforms through pattern recognition, making them ideal candidates for deep learning approaches. Physics-based tasks, such as earthquake location and focal mechanism determination, require precise physical constraints in travel-time and first-motion polarity, making them well-suited for geophysical inversions. 
With accurate and complete measurements of seismic waveforms from the deep learning-based perception tasks, geophysical inversions can efficiently determine precise earthquake source parameters.
The distinction of perception-based and physics-based tasks here is based in the common methods employed for these tasks, rather than a strict definition. For example, phase association could be solved through multiple approaches based on waveform features or travel-time constraints \citep{mcbrearty2019pairwise,munchmeyer2024pyocto}.
Our multitask model is designed to address perception-based tasks of earthquake detection and phase picking and provide constraints (i.e., the origin time) for phase association, all within an end-to-end process. The extracted information enables subsequent estimation and refinement of earthquake locations using dedicated algorithms such as NonLinLoc \citep{lomax2000probabilistic} and HypoDD \citep{waldhauser2000double} and focal mechanism determination using HASH \citep{hardebeck2003using} to construct comprehensive earthquake catalogs.
This integrated approach combines the efficiency of deep learning for pattern recognition with the precision of physics-based inversions, enabling robust and automated earthquake monitoring and efficient seismic data mining.

\subsection{Training details}
\paragraph{Training datasets} We utilized the California Earthquake Event Dataset (CEED) for training PhaseNet+. CEED is a comprehensive dataset containing seismic waveforms, phase arrival-time picks, first-motion polarities, ground motion intensities, and earthquake source information from California \citep{zhu2025california}. This comprehensive collection of labels makes it ideal for training our multitask model with the necessary information for phase picking, polarity picking, and event detection. The data from North California was used during training, containing 325K events and 1.1M three-component waveforms with corresponding 1.1M pairs of P and S picks, 1.0M up and down polarity picks. Each waveform has 12,000 samples with a sampling rate of 100Hz. During training, a randomly selected segment of length 4,096 is extracted as input. The model is trained with the AdamW \citep{loshchilov2017decoupled} optimizer using a batch size of 256 for 100 epochs. The learning rate is set to 0.001 and follows a schedule with a one-epoch linear warmup and the cosine decay strategy.

\section{Results}

To evaluate the performance of the multitask model, we first benchmark each task model's performance on the test dataset, then demonstrate the multitask model's end-to-end performance using a real-world earthquake sequence. 

\subsection{Test dataset evaluation}
\subsubsection{Arrival-time picking}
\begin{figure}
    \begin{subfigure}{\textwidth}
    \centering
    \includegraphics[width=0.7\linewidth]{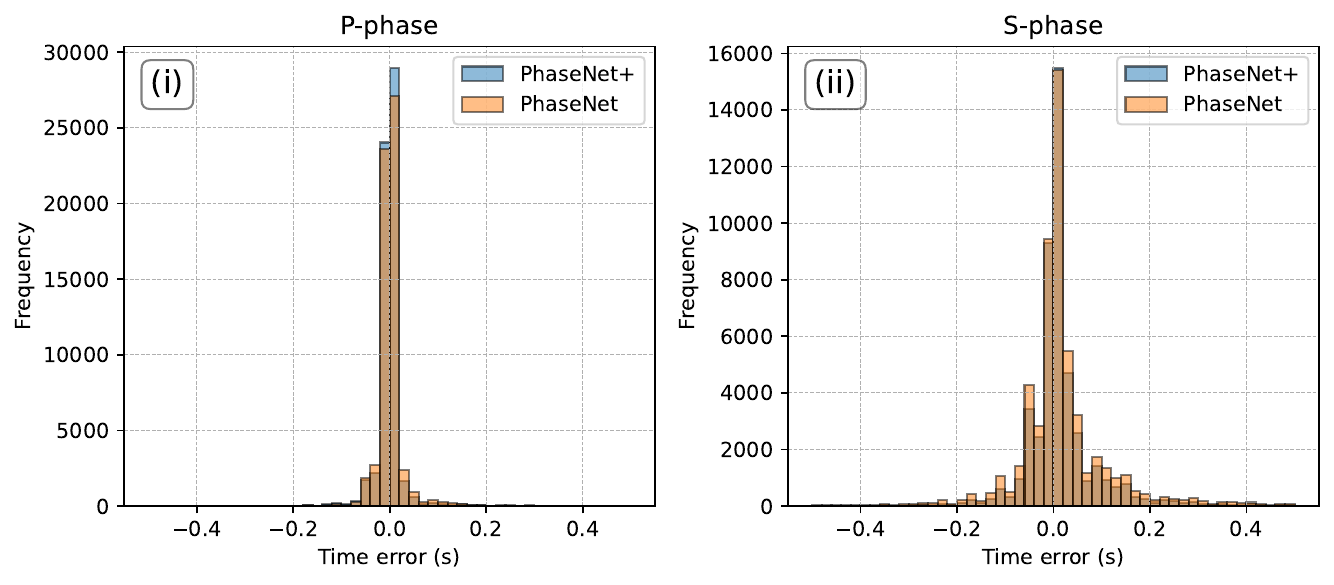}
    \caption{}
    \end{subfigure}
    \begin{subfigure}{\textwidth}
    \centering
    \includegraphics[width=0.7\linewidth]{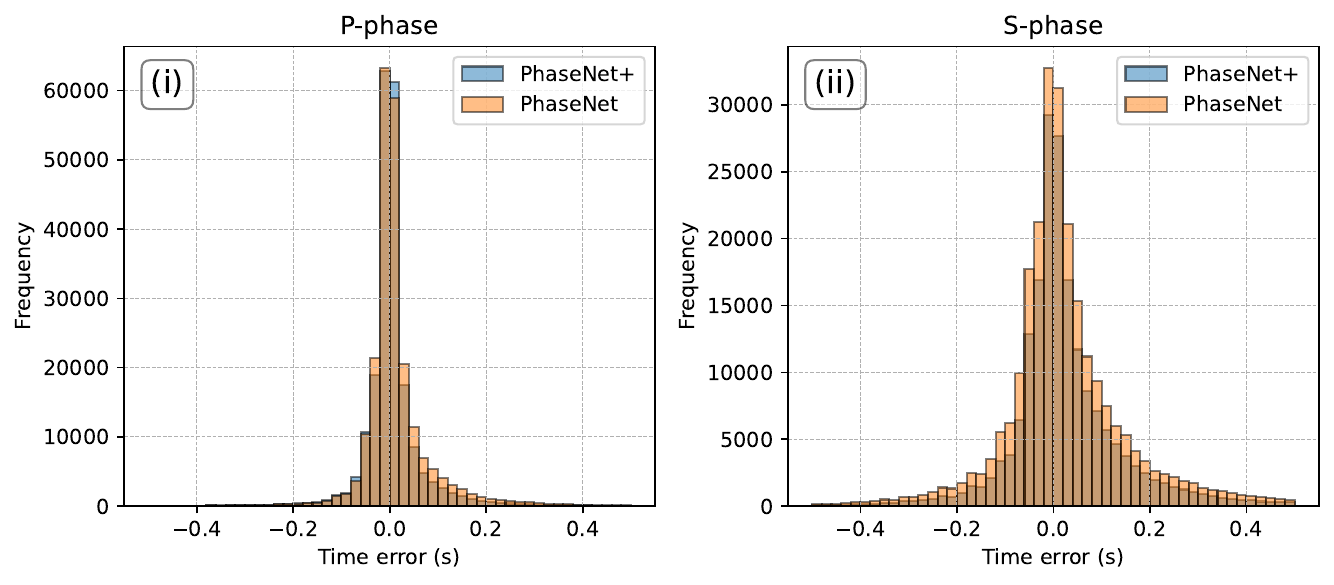}
    \caption{}
    \end{subfigure}
    \caption{Comparison of phase arrival-time picking performance between PhaseNet+ and PhaseNet models evaluated on (a) Northern and (b) Southern California datasets.}
    \label{fig:time_error}
\end{figure}

Since the PhaseNet+ model builds upon the original PhaseNet architecture, similar performance is expected in phase arrival-time picking performance. \Cref{fig:time_error} compares the phase arrival-time picking performance between the PhaseNet+ and original PhaseNet models. Both models were trained on the Northern California dataset before 2023 and applied to Northern and Southern California datasets from 2023. The comparable phase picking performance between the two models indicates that jointly training the three tasks does not compromise the phase arrival-time picking accuracy, but also does not yield further improvements.

\subsubsection{Polarity Picking}

\begin{figure}
    \centering
    \includegraphics[width=\linewidth]{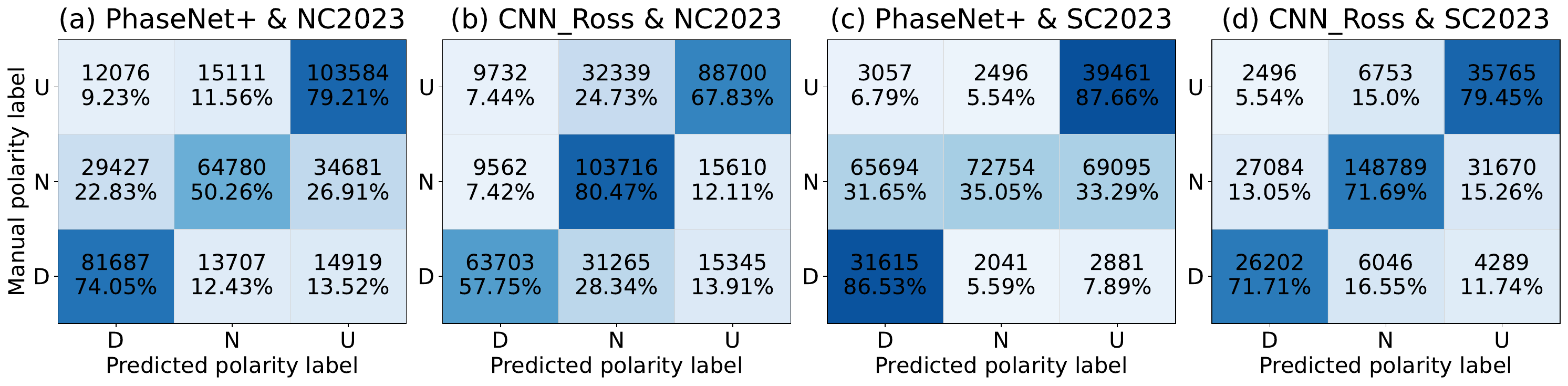}
    \caption{Comparison of polarity picking performance between PhaseNet+ and the CNN\_Ross model, evaluated on Northern California (NC2023) and Southern California (SC2023) datasets.}
    \label{fig:performance_comparison}
\end{figure}

\begin{figure}
    \centering\includegraphics[width=0.7\linewidth]{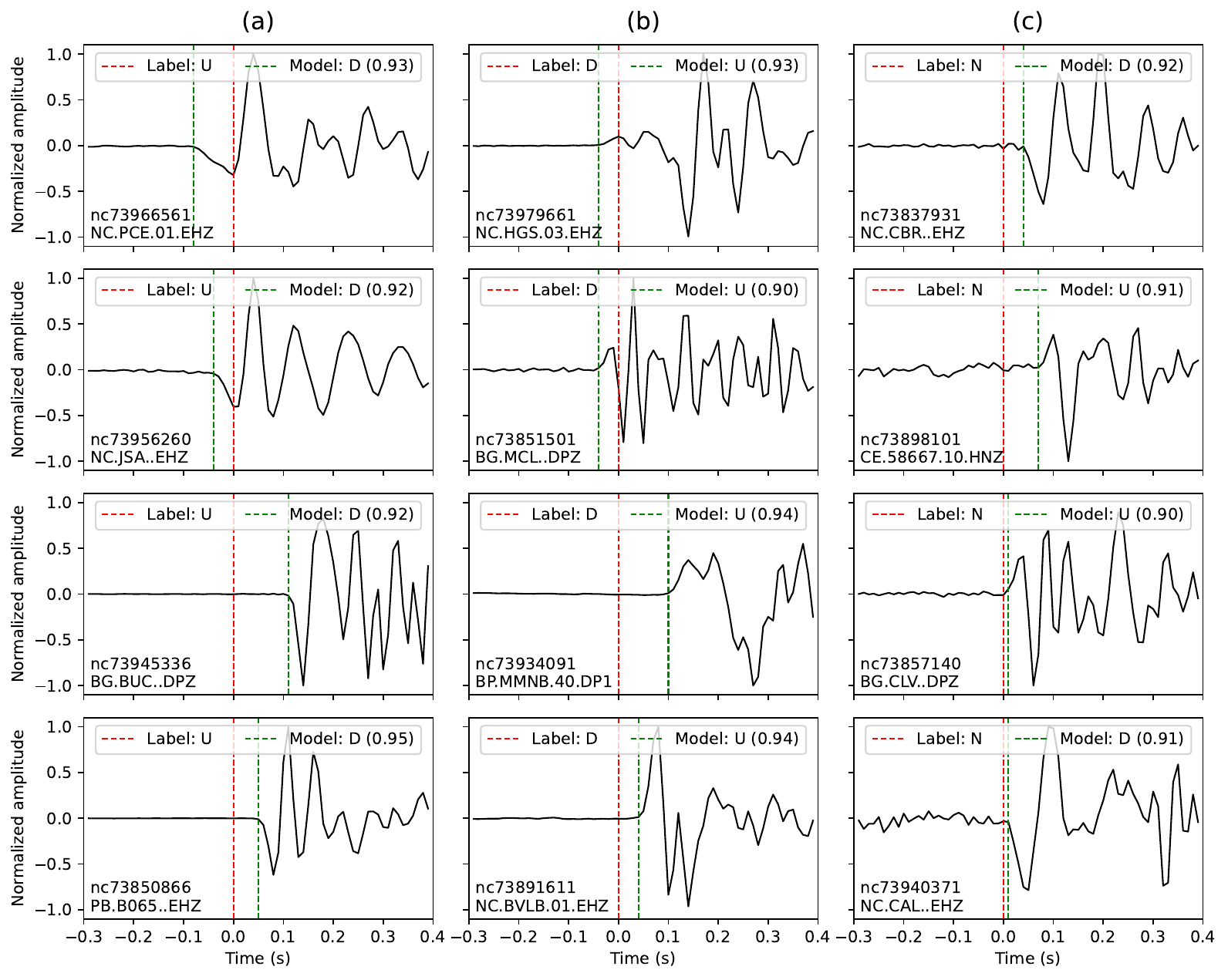}
    \caption{Examples of mismatched polarities between manual labels and PhaseNet+ model predictions: (a) and (b) Opposite polarities arise from differing interpretations of phase arrival times. (c) Cases where polarity labels are marked as unknown (``N") but are classified by PhaseNet+. All the waveforms are 0.3 s prior to and 0.4 s after the manually picked P arrivals.}
    \label{fig:polarity_error}
\end{figure}

We evaluated PhaseNet+'s polarity picking performance in comparison to the state-of-the-art convolutional neural network model developed by \citet{ross2018p} (referred to as CNN\_Ross). The CNN\_Ross is pre-trained on earthquake data from Southern California prior to 2018. Both models were applied to the 2023 test dataset from Northern and Southern California, containing impulsive and emergent P arrivals with manually assigned polarities: upward (``U"), downward (``D"), or unknown (``N"). To maintain consistency with the polarity labeling schemes, we categorized PhaseNet+ polarity predictions as follows: values below -0.33 as ``D", between -0.33 and 0.33 as ``N", and above 0.33 as ``U". \Cref{fig:performance_comparison} demonstrates that PhaseNet+ achieves comparable or superior performance to CNN\_Ross for both upward and downward polarities. 
Furthermore, PhaseNet+ successfully reclassified approximately 50\%-60\% of unknown polarity (``N") cases into ``U" or ``D" categories. Many of these reclassified examples exhibit high signal-to-noise ratios and clear polarities (\Cref{fig:polarity_error}b), suggesting that PhaseNet+ could increase the number of usable first-motion polarities and enhance focal mechanism inversion.
To investigate the $\sim$10\%-20\% discrepancy between manual labels and model predictions in \Cref{fig:performance_comparison}, we examined the corresponding waveforms with ambiguous polarities in manual labels. \Cref{fig:polarity_error}a shows cases where model predictions differ from manual labels primarily due to adjustments in selected phase arrival-times affecting polarity determination, with PhaseNet+ often providing more reasonable predictions in these challenging scenarios.

\subsubsection{Origin time prediction for phase association} 

\begin{figure}
    \begin{subfigure}{\textwidth}
        \centering
        \includegraphics[width=0.7\linewidth]{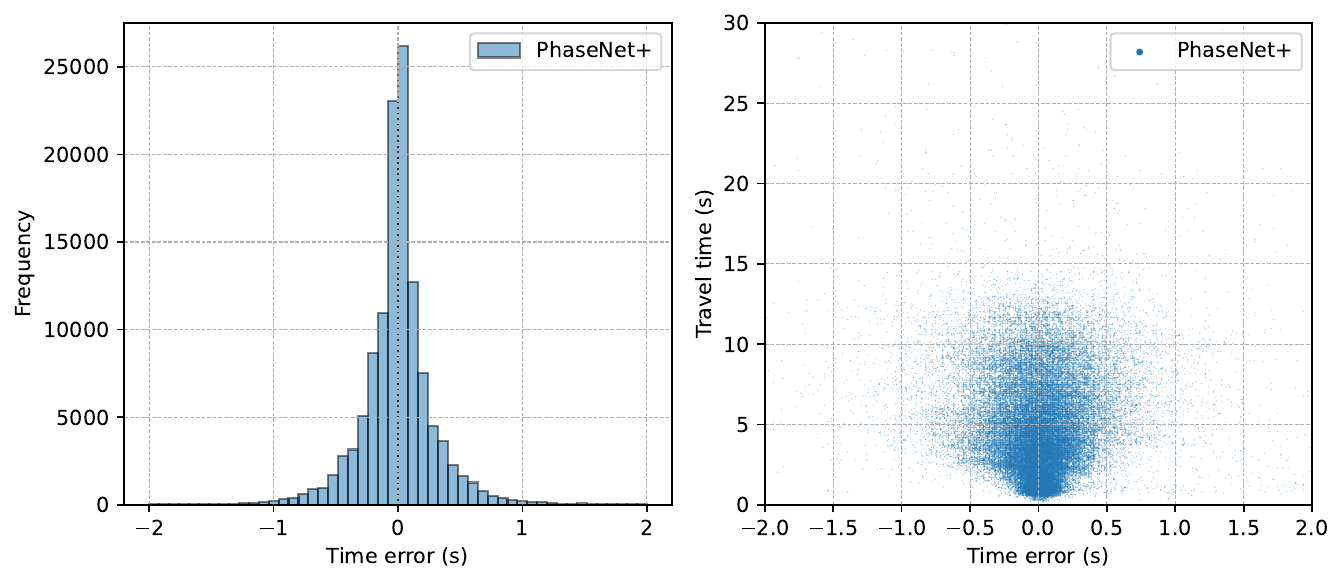}
        \caption{}
    \end{subfigure}
    \begin{subfigure}{\textwidth}
        \centering
        \includegraphics[width=0.7\linewidth]{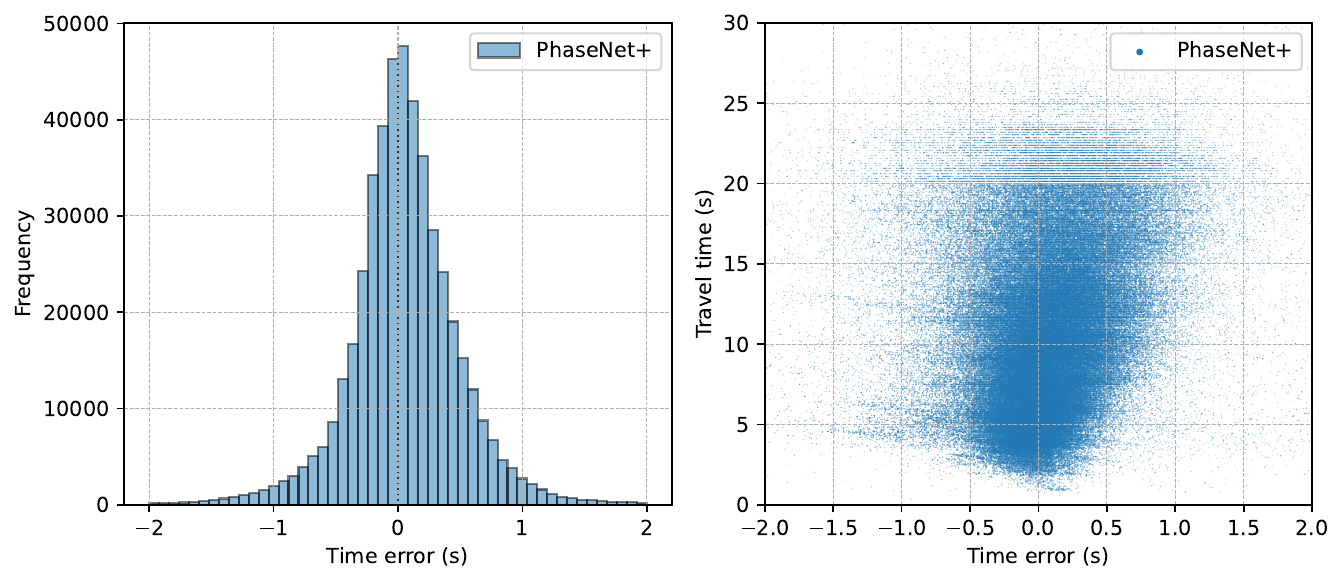}
        \caption{}
    \end{subfigure}
    \caption{Origin time prediction performance of PhaseNet+ evaluated on (a) Northern and (b) Southern California datasets.}
    \label{fig:event_time_error}
\end{figure}

PhaseNet+ incorporates origin time prediction as a novel task to facilitate phase association, so we need to validate its prediction accuracy to ensure reliable constraints for phase association. 
Although precise origin time needs to be determined through geophysical inversion during earthquake location, approximate origin time predictions are sufficient for phase association purposes. 
\Cref{fig:event_time_error} shows the origin time error distributions and the correlation with phase travel times. 
For both Northern and Southern California, prediction errors remain relatively stable within 2s, with a modest increase at longer distances (e.g., travel times exceeding 15s). This is expected due to stronger impacts of velocity model inaccuracies and limited large events in model training. 
The origin time errors are smaller than typical earthquake intervals of enhanced earthquake catalogs, such as the QTM catalog, which has an average inter-event time of 174s and merged detections separated by less than 2s during template matching \citep{ross2019searching}. Therefore, the origin time prediction task could be effectively applied to constrain the identification of coherent phase arrivals from the same seismic event, thereby streamlining the phase association process.
Additionally, the comparable performance between Northern and Southern California datasets underscores the applicability of PhaseNet+'s origin time predictions across various seismic network configurations and geological settings.

\subsection{Application to the 2019 Ridgecrest Earthquake Sequence}

The evaluation based on the test dataset validates the multitask model's performance in phase arrival-time picking, phase polarity picking, and origin time prediction for phase association on preselected waveforms.
To demonstrate the model's end-to-end performance, we applied PhaseNet+ to the 2019 Ridgecrest earthquake sequence following the workflow in \Cref{fig:workflow}.
The 2019 Ridgecrest earthquake sequence comprised a Mw 6.4 foreshock on July 4, followed by a Mw 7.1 mainshock approximately 34 hours later and extensive subsequent aftershocks that revealed a complex fault network at depth \citep{ross2019hierarchical,shelly2020high}. 
The Southern California Seismic Network (SCSN) routinely cataloged earthquakes and focal mechanisms as the sequence progressed. 
Several single-task deep learning models \citep{ross2018p,zhu2019phasenet} demonstrated enhanced performance in rapidly developing and enhancing aftershock catalogs, facilitating detailed analysis of seismicity evolution, fault structures, and spatiotemporal variations in stress conditions \citep{liu2020rapid, cheng2020variations}. 
To demonstrate the effectiveness of our multitask model PhaseNet+, we tested it on the initial phase of the 2019 Ridgecrest earthquake sequence and compared our results with existing catalogs.

\begin{figure}
    \centering
    \includegraphics[width=0.8\linewidth]{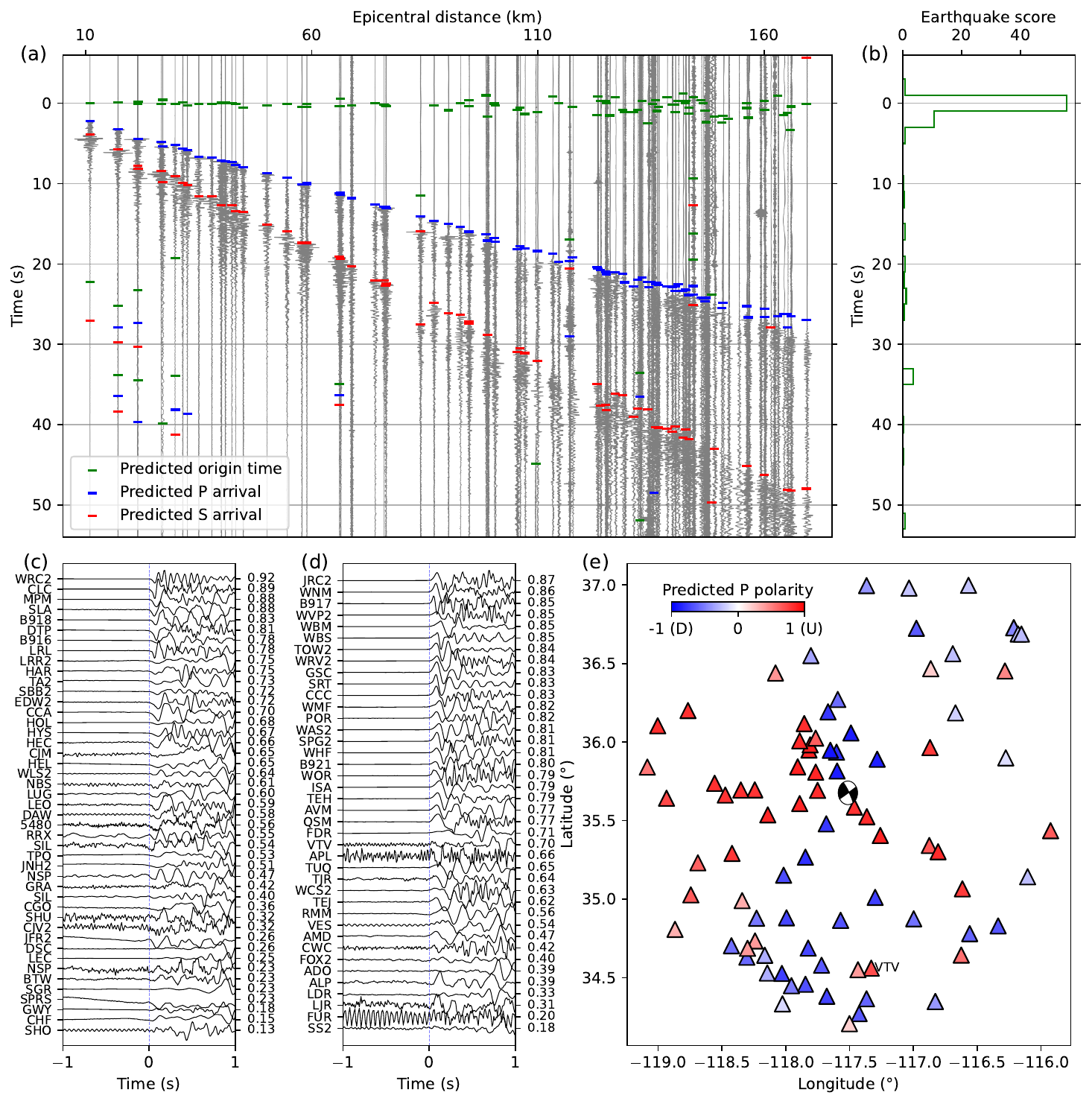}
    \caption{Example of PhaseNet+ predictions for an event from the Ridgecrest sequence (event ID: ci38445015). (a) Vertical component waveforms sorted by the epicentral distance, with blue and red ticks marking PhaseNet+ P and S picks, and green ticks indicating predicted earthquake origin times. (b) Histogram of predicted earthquake origin times aggregated across stations, showing prediction consistency. Vertical component waveforms within 2-second windows centered on P arrivals, shown alongside predicted (c) negative and (d) positive first-motion polarity scores. (e) The spatial distribution of stations with P picks associated. Stations are color-coded by the predicted first-motion polarity scores. The beachball shows the location and focal mechanism solution obtained by SCSN.}
    \label{fig:multipredictions}
\end{figure}

We downloaded one week (July 4-10, 2019) of continuous seismic waveforms from stations with network codes ``CI", ``PB", ``NN", ``NP", and ``SN" within 1.5 degrees of the coordinates (117.504°W, 35.705°N). This time period includes the largest foreshock, the mainshock, and several days of the dense aftershock sequence. At each station, we selected data from a single instrument type following the priority order: HH$>$EH$>$HN$>$BH. We applied PhaseNet+ to pick phase arrivals and polarities, and predict event origin times for association (\Cref{fig:multipredictions}).
Event locations were determined using ADLoc \citep{zhu2025robust} with the 1D layered velocity model of SCSN catalog \citep{hutton2010earthquake}. For comparison, we generated a separate catalog using GaMMA \citep{zhu2022earthquake} for phase association based on the PhaseNet+ pick times, but without using the predicted origin times.
We use ADLoc for location in this reference catalog as well. We refer to these two catalogs as the PhaseNet+ and GaMMA catalogs, respectively.

\begin{figure}
    \centering
    \includegraphics[width=0.8\linewidth]{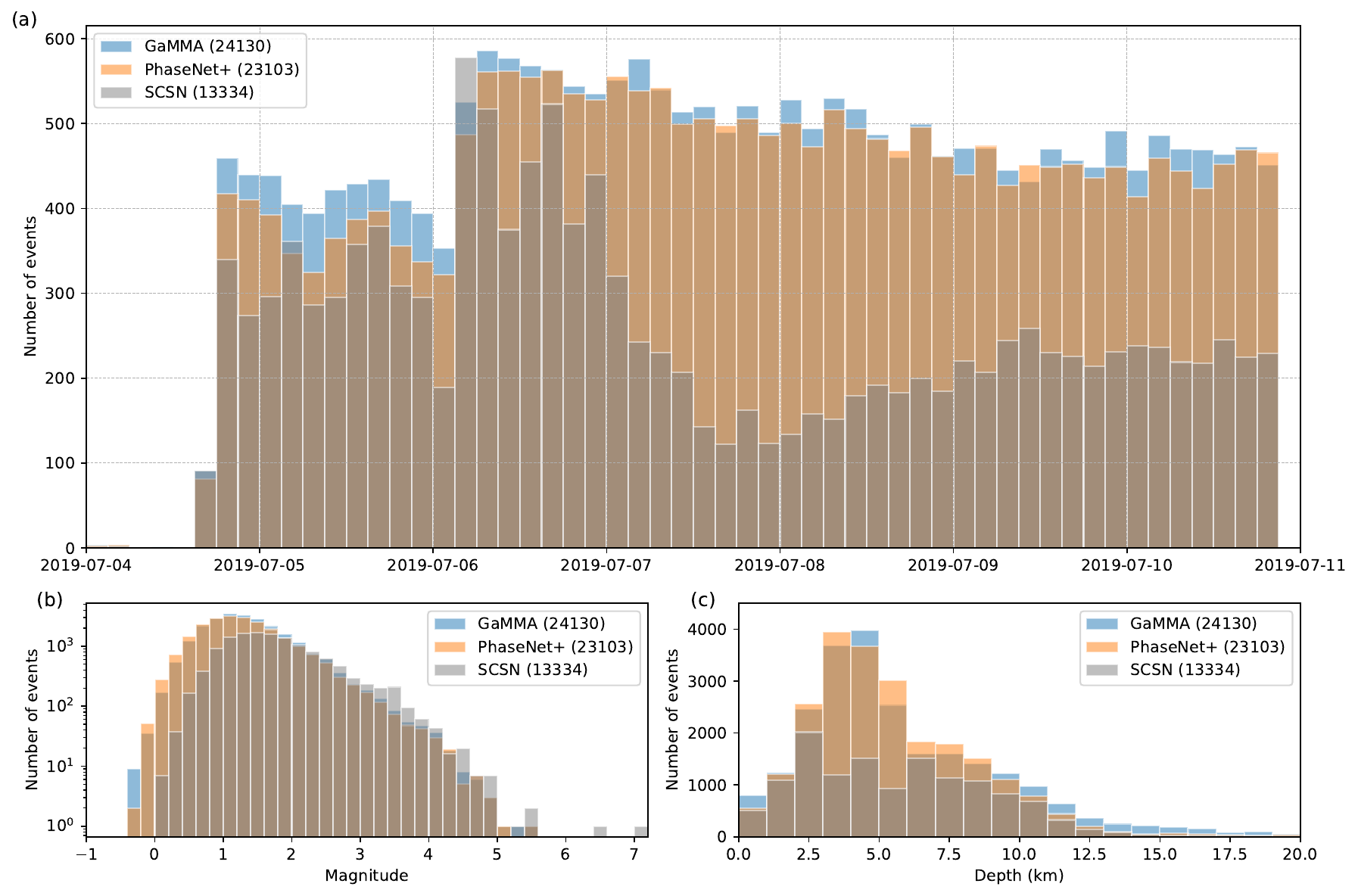}
    \caption{Comparison of (a) temporal distribution of event frequency, (b) magnitude distribution, and (c) depth distribution across different catalogs.}
    \label{fig:event_time_mag}
\end{figure}

\begin{figure}
    \centering
    \includegraphics[width=\linewidth]{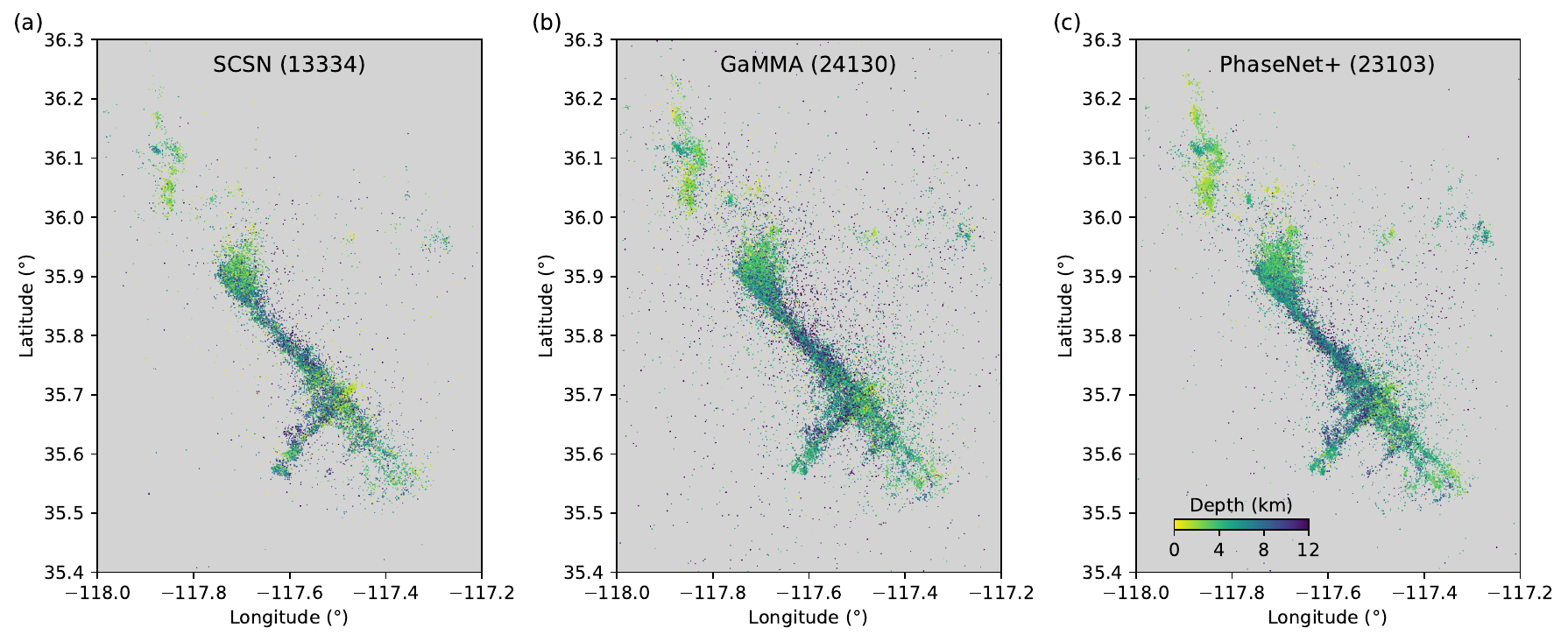}
    \caption{Spatial distribution of earthquakes from (a) SCSN catalog, (b) GaMMA catalog, and (c) PhaseNet+ catalog.}
    \label{fig:event_map}
\end{figure}

The earthquake frequency and magnitude distributions of the PhaseNet+ and GaMMA catalogs are overall similar, with each catalog containing 73\% to 81\% more events than the SCSN catalog (\Cref{fig:event_time_mag}a-b). The depth distribution of both catalogs remains consistent with the SCSN catalog, with most newly detected events occurring between 3 and 6 km depth (\Cref{fig:event_time_mag}c), coinciding with the depth range of interpreted orthogonal subfaults \citep{ross2019hierarchical}. The three catalogs follow similar trends in the first two days of the sequence, as the SCSN catalog was carefully manually reviewed during this period, leading to a higher number of events and more complete documentation. This also explains the decrease in earthquake numbers in the following days in the SCSN catalog.
Although the PhaseNet+ catalog contains 4\% fewer events than the GaMMA catalog, its workflow is more efficient (\Cref{fig:workflow}), and the resulting catalog exhibits less spatial scatter (\Cref{fig:event_map}).
This suggests a slightly higher number of false-positive detections in the GaMMA catalog.
Notably, the fault branches and orthogonal fault structures at the southeast end are more clearly delineated (\Cref{fig:event_map}c). These results demonstrate that for complex earthquake sequences, rapid association based on predicted event origin times works as effectively for dense aftershocks as dedicated phase association algorithms, making the end-to-end approach using a multitask model rival state-of-the-art methods.

Moreover, the polarities simultaneously predicted by PhaseNet+ enable the determination of focal mechanisms to provide constraints on faulting style and stress field orientation. We applied the SKHASH package \citep{skoumal2024skhash} to determine focal mechanism solutions, utilizing the absolute polarity scores between 0 and 1 from PhaseNet+ as weights of first-motion polarity quality (\Cref{fig:multipredictions}). We selected only polarities with absolute scores above 0.1 and events with at least 8 polarity measurements and maximum epicenter distance of 120 km for focal mechanism inversion.
We determined two sets of focal mechanism solutions using either polarity data alone or both polarity and S\/P amplitude ratio data.

\begin{figure}
    \centering
    \includegraphics[width=0.8\linewidth]{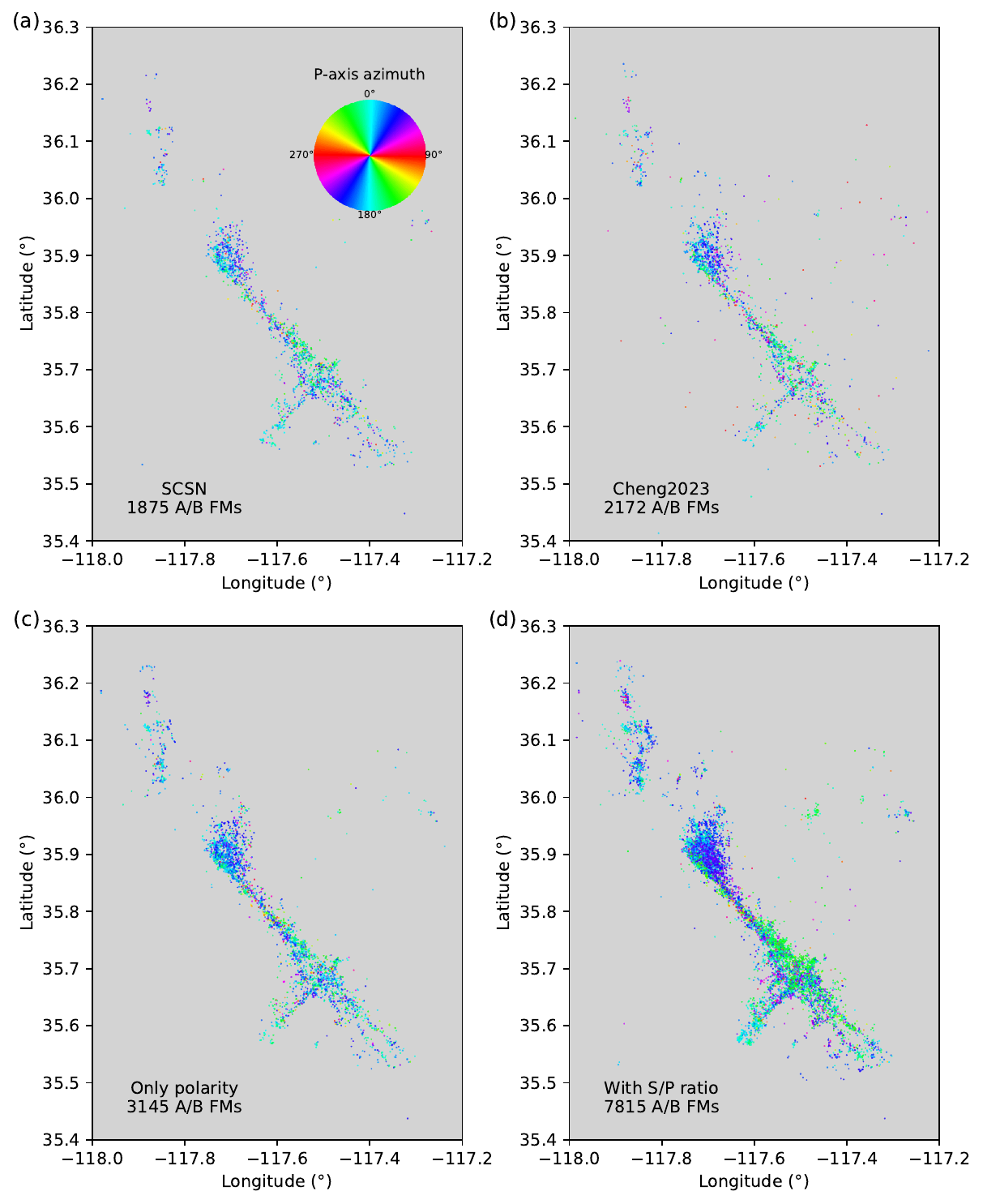}
    \caption{Spatial distribution of P-axis azimuths from focal mechanism solutions: (a) SCSN catalog, (b) Cheng2023 catalog, (c) PhaseNet+ catalog using only polarity data, and (d) PhaseNet+ catalog using both polarity and S\/P amplitude ratio data.}
    \label{fig:p_trend}
    \end{figure}

\begin{figure}
    \centering
    \includegraphics[width=0.8\linewidth]{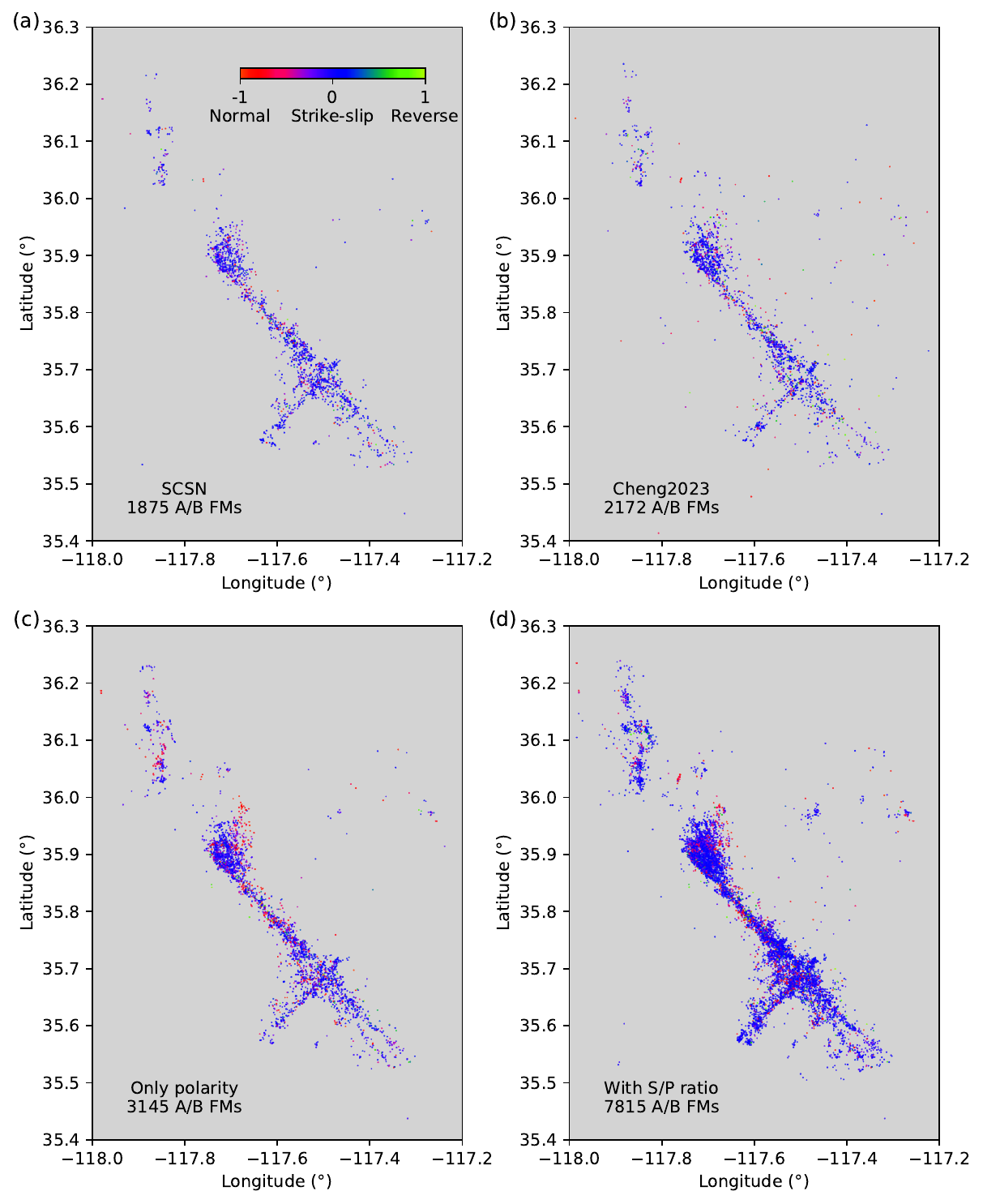}
    \caption{Spatial distribution of scalar faulting types derived from focal mechanism solutions: (a) SCSN catalog, (b) Cheng2023 catalog, (c) PhaseNet+ catalog using only polarity data, and (d) PhaseNet+ catalog using both polarity and S\/P amplitude ratio data. }
    \label{fig:fm_type}
\end{figure}

We compared the results of quality A and B focal mechanisms with two existing catalogs: the SCSN catalog \citep{yang2012computing} and \citet{cheng2023refineda}'s enhanced catalog. Both catalogs used the HASH algorithm \citep{hardebeck2002new} based on polarity picks and S/P amplitude ratios. \citet{cheng2023refineda} applied the CNN\_Ross model to detect additional phases and polarities, complementing SCSN measurements to enhance focal mechanism solutions.
The revealed faulting characteristics are consistent across catalogs (\Cref{fig:p_trend,fig:fm_type}), both on the main fault and on minor substrands. The number of focal mechanism solutions increased up to about four-fold: from 1,875 in the SCSN catalog to 2,172 in \citet{cheng2023refineda}'s catalog, 3,145 using only polarity data from PhaseNet+, and 7,815 using both polarity and S/P amplitude ratios. \Cref{fig:p_trend} shows that this enhanced dataset better illuminates the spatial variations in P-axis azimuth, often considered a proxy for maximum horizontal compressive stress direction. The P-axis directions lie approximately in the N-S direction, with clear orientations from the middle to the northwestern rupture terminus, consistent with previous observations \citep{cheng2020variations, wang2020seismotectonics}. Following \citet{shearer2006comprehensive}, we also parameterized focal mechanism types into scalars representing faulting types ranging from -1 (normal) to 0 (strike-slip) to 1 (reverse), as shown in \Cref{fig:fm_type}. Results across four catalogs consistently indicate the predominance of strike-slip events throughout the sequence with some normal faulting events. The enhanced focal mechanism catalog confirms previous observations of the regional stress field and reveals finer-scale variations in faulting styles and stress orientations.

\section{Discussion}

Deep learning offers a significant advantage by enabling end-to-end learning approaches, eliminating the need for traditional feature engineering steps and allowing direct learning from raw data.
In addition to eliminating feature engineering for individual tasks, many research problems have further integrated multi-step processing workflows into unified end-to-end learning frameworks, substantially improving both efficiency and accuracy across various applications.
For example, end-to-end deep learning models have become the standard approach in speech recognition and generation, directly processing raw audio waveforms to generate text transcriptions or vice versa \citep{amodei2016deep,chen2024f5tts}.
These end-to-end solutions not only streamline workflows but also enhance overall performance by eliminating intermediate steps that could introduce errors or information loss.
In earthquake monitoring, the development of end-to-end frameworks remains an active area of research \citep{zhang2023simultaneous,si2024all,zhu2022end}.
Our previous work, EQNet, explored an end-to-end solution using multiple stations but did not incorporate polarity picking \citep{zhu2022end}. This study advances single-station processing by integrating phase picking, polarity picking, and event detection into a unified model. This framework can be readily extended to multi-station scenarios by incorporating multi-station processing capabilities, such as neural operators \citep{sun2023phase}.

The multitask model builds upon the success of deep learning in phase picking, polarity picking, and earthquake location. It simplifies the earthquake monitoring workflow and facilitates the development of enhanced earthquake catalogs, lowering the magnitude of completeness through effective detection of small earthquakes and increasing the quantity and quality of focal mechanism solutions through reliable determinations of phase polarities.
Enhanced earthquake catalogs enable detailed imaging of fault structures and comprehensive analysis of spatial-temporal seismic activity patterns, while refined focal mechanism solutions enable more precise characterization of fault slip and stress orientations. These improvements contribute to a more comprehensive understanding of earthquake faulting processes, offering valuable insights for seismic hazard assessment and tectonic processes. 

Despite the advantages of the multitask model, several challenges remain.
First, multitask learning requires optimization of combined losses from all tasks, necessitating careful balance between loss weights and learning rates. In this study, we applied a weight of 0.2 to the polarity loss while maintaining other weights at 1.0 to make all losses converge to similar scales. Future work could implement more advanced techniques, such as dynamic weighting and learning rate scheduling, to further optimize model performance \citep{guo2018dynamic,liu2019end}.
Second, the multitask model requires training datasets with sufficient labels for all included tasks, potentially limiting its application with existing datasets. In this work, we utilized the CEED dataset, which provides comprehensive labels for phase arrival times, phase polarities, and event origin times. Future work could explore semi-supervised learning approaches to address missing labels and leverage additional datasets \citep{zhu2022introduction,zhu2023seismic}.
Last, our current approach to phase association relies on origin time prediction, which becomes less effective when multiple earthquakes occur nearly simultaneously, e.g., in dense aftershock sequences.
To address this limitation, one direct solution is to incorporate station locations or waveform similarity metrics to differentiate events, enabling distinction between events with similar origin times but different locations or waveform characteristics.
Another more advanced approach would be to expand the multitask model to include source location prediction to enhance phase association performance. These limitations could be addressed in future research to further improve the model's performance and broaden its applicability.

\section{Conclusion}

The development of integrated, efficient, and accurate frameworks for seismic data analysis is essential for advancing earthquake monitoring and characterization. In this study, we introduced PhaseNet+, a multitask deep learning framework designed to simultaneously perform phase arrival-time picking, first-motion polarity determination, and origin time prediction for phase association. This integrated approach provides a simplified and efficient end-to-end solution for earthquake monitoring workflows.
The evaluation of PhaseNet+ demonstrated robust performance across all tasks on test datasets, achieving comparable or superior accuracy relative to existing state-of-the-art single-task models. The application to the 2019 Ridgecrest earthquake sequence further highlights its capability to efficiently develop enhanced earthquake catalogs with more high-quality focal mechanism solutions, illuminating fault structures, stress fields, and earthquake source processes.
The proposed multitask framework extends beyond the specific implementation of PhaseNet+, serving as a template for developing integrated deep learning models to provide an efficient and reliable approach for generating comprehensive earthquake catalogs offering valuable insights into earthquake processes.

\section*{Open research statement}
The implementation of PhaseNet+ and the associated training scripts is available at \url{https://github.com/AI4EPS/EQNet}.
PhaseNet+ will be integrated into the SeisBench \citep{woollam2022seisbench} framework at latest by the time of acceptance of this manuscript.

\section*{Acknowledgements}
This research was supported by the United States Geological Survey (USGS) under Grant No. G25AP00119.
Waveform data, metadata or data products for this study were accessed through the Northern California Earthquake Data Center \citep{ncedc2014northern} and Southern California Earthquake Data Center \citep{scedc2013southern}.
We thank Robert Skoumal for valuable discussions and suggestions on running SKHASH for inverting focal mechanisms \citep{skoumal2024skhash}.

\newpage
\clearpage
\bibliography{references}

\end{document}